\documentclass[prb,preprint]{revtex4}

\pdfoutput=1

\usepackage{graphicx}

\begin{document}

\title{Comment on the thermal Hall effect in cuprates}


\medskip 

\date{January 3, 2022} \bigskip

\author{Manfred Bucher \\}
\affiliation{\text{\textnormal{Physics Department, California State University,}} \textnormal{Fresno,}
\textnormal{Fresno, California 93740-8031} \\}

\begin{abstract}

Insights from stripe incommensurabilities and antiferromagnetic stability indicate that the magnetic moments of both host $Cu^{2+}$ ions and $Cu$ atoms from electron doping support the thermal Hall effect in cuprates, whereas those of $O$ atoms from hole doping oppose it.

\end{abstract}

\maketitle
A thermal Hall effect—a temperature difference transverse to a heat flow in a perpendicular magnetic field—was recently discovered in a variety of copper oxides, including hole-doped and electron-doped high-$T_c$ superconductors, their undoped parent crystals (being Mott insulators), but also the antiferromagnet $Cu_3TeO_6$ which, lacking a half-filled band, is \textit{not} a Mott insulator.\cite{1,2,3,4,5} By systematic experimentation it has been established that the dominant heat transport is carried by phonons. Other results indicate that the mechanism of the heat-flow chirality arises from a combination of scattering by defects and short-range antiferromagnetic correlations.\cite{5} Details of the corresponding mechanism and the kind of responsible defects remain open questions. 
What is conspicuously absent, though, is a consideration of the role of the \textit{magnetic moments}, in their respective modifications, of the two elements common to all cases—\textit{copper} and \textit{oxygen}. The aim of this note is to bring attention to them.

An analysis of stripe incommensurability\cite{6} has revealed that doped holes in the $CuO_2$ planes reside pairwise at oxygen atoms, $2e^+ + O^{2-} \rightarrow O$, whereas doped electrons reside reside pairwise at copper atoms, $2e^- + Cu^{2+} \rightarrow Cu$. 
The influence of their magnetic moments provides a qualitative explanation for the narrow antiferromagnetic (AFM) phase of hole-doped lanthanide cuprates (N\'{e}el point $p_0^N = 0.02$) but a wide AFM phase of the electron-doped compounds ($p_0^N \approx 0.13$). 
In the hole-doped crystals, $\mathbf{m}(O)$ moments of the defect $O\;2p^6$ atoms on anion lattice sites in the $CuO_2$ plane and with spin quantum number $s=1$ are interspersed with the $\mathbf{m}(Cu^{2+})$ moments of the host crystal on cation lattice sites and of spin quantum number $s=\frac{1}{2}$.\cite{6}
The different sites and strengths of the magnetic moments, $m(O) \approx 2 m(Cu^{2+})$,
strongly upset 3D-AFM of the host and cause its collapse at $p_0^N = 0.02$.
In the electron-doped crystals, by contrast, the $\mathbf{m}(Cu) \simeq \mathbf{m}(Cu^{2+})$ moments of the defect $Cu\;3d^{10}4s^1$ atoms and host $Cu^{2+}\;3d^{9}$ ions, both on cation lattice sites and with comparable strength, account for the wide AFM phase of the electron-doped compounds.\cite{6}

It seems therefore reasonable to compare the thermal-conductivity experiments on hole-doped and electron-doped lanthanide cuprates\cite{1,5} in light of the extant magnetic moments. In both families the regular thermal conductivity $\kappa_{xx}$ decreases with increasing doping.
Contrarily, the thermal \textit{Hall} conductivity $\kappa_{xy}$ decreases with hole doping but increases with electron doping.\cite{1,5}
Similar to the case of AFM stability, the interspersing of the host
$\mathbf{m}(Cu^{2+})$ moments at cation lattice sites by the defect $\mathbf{m}(O)$ moments on anion lattice sites may account for the weakening of the Hall thermal conductivity $\kappa_{xy}$ with hole doping, whereas the (same-site) replacement of host-crystal $\mathbf{m}(Cu^{2+})$ moments by the comparable $\mathbf{m}(Cu)$ moments may account for the increase of $\kappa_{xy}$ with electron doping.
If the analogy with AFM stability is valid, $\mathbf{m}(Cu^{2+})$ moments would support the thermal Hall effect—and $\mathbf{m}(Cu)$ moments even more—but $\mathbf{m}(O)$ moments would oppose it.

Qualitatively, the increase of of $\kappa_{xy}$ with electron doping (via rare-earth replacement, $Ln^{3+} = La^{3+}, Pr^{3+}, Nd^{3+} \rightarrow  Ce^{4+}$),\cite{6} has raised the question as to the role of $Ce^{4+}$ ions, residing in the sandwiching $[Ln/Ce]O$ layers, as possible skew scatterers in electron-doped compounds.\cite{5} The question is intimately related to the defect $Cu$ atoms in the $CuO_2$ plane.
Thus, skew scattering may arise from $Ce^{4+}$ or $Cu$ or both.

No thermal Hall conductivity is observed for hole-doped cuprates beyond the quantum critical point, $x > x^*$, that is, for the overdoped compounds outside the pseudogap phase.\cite{3}  
Because of good metalicity,\cite{7} the thermal current is carried here exclusively by free charges (Wiedemann-Franz law),\cite{1} not by phonons—hence no skew scattering of phonons in this doping range.

\end{document}